\documentclass{raa}
\usepackage{graphicx, times}
\usepackage{natbib}
\usepackage{amssymb, amsmath}
\bibpunct{(}{)}{;}{a}{}{,}
\def\degree{\hbox{$^\circ$}}

\usepackage[a4paper=true, pagebackref=true]{hyperref}
\hypersetup{pdftitle = The title of my PDF, pdfauthor = My name, pdfsubject= The subject, pdfkeywords = keyword1 keyword2 keyword3}
\hypersetup{colorlinks = true, linkcolor = green, anchorcolor = red, citecolor = blue, filecolor = red, pagecolor = red, urlcolor = red}
\begin{document}

\title{A polarization study of the supernova remnant CTB~80\footnote{\small Supported by the National Natural Science Foundation of China.}}

\volnopage{{\bf 202X} Vol.\ {\bf X} No. {\bf XX}, 000--000}
\setcounter{page}{1}

\author{Xianghua Li\inst{1}, Xiaohui Sun\inst{1}, Wolfgang Reich\inst{2}, Xuyang Gao\inst{3,4}}
\institute{ Department of Astronomy, Yunnan University, and Key Laboratory of Astroparticle Physics of Yunnan Province, Kunming 650091, China; {\it xhli@ynu.edu.cn, xhsun@ynu.edu.cn}\\
\and
 Max-Planck-Institut f\"ur Radioastronomie, Auf dem H\"ugel 69, 53121 Bonn, Germany\\
\and
National Astronomical Observatories, CAS, Jia-20 Datun Road, Chaoyang District, Beijing 100101, China\\
\and
CAS Key Laboratory of FAST, National Astronomical Observatories, Chinese Academy of Sciences\\
\vs \no
{\small Received 202X Month Day; accepted 202X Month Day}
}

\abstract{
We present a radio polarization study of the supernova remnant CTB~80 based on images at 1420~MHz from the Canadian Galactic plane survey, at 2695~MHz from the Effelsberg survey of the Galactic plane, and at 4800~MHz from the Sino-German $\lambda$6~cm polarization survey of the Galactic plane. We obtained a rotation measure (RM) map using polarization angles at 2695~MHz and 4800~MHz as the polarization percentages are similar at these two frequencies. RM exhibits a transition from positive values to negative values along one of the shells hosting the pulsar PSR B1951+32 and its pulsar wind nebula. The reason for the change of sign remains unclear. We identified a partial shell structure, which is bright in polarized intensity but weak in total intensity. This structure could be part of CTB~80 or part of a new supernova remnant unrelated to CTB~80.
}

\authorrunning{X. Li et al.}
\titlerunning{Polarization study of CTB~80}
\maketitle

\section{Introduction}

CTB~80 is a prominent radio source with a complex morphology consisting of a bright central source embedded in a plateau and extended ridges or shells. \citet{velusamy+74} suggested that CTB~80 is a supernova remnant (SNR) based on its non-thermal spectrum and high polarization percentage of 15 -- 20\% at $\lambda$11~cm. \citet{asv+81} confirmed that CTB~80 is an SNR from high-resolution observations and proposed that the central source probably is a Crab-like SNR, because it is highly polarized and its spectrum is flat. Follow-up X-ray~\citep{becker82} and radio~\citep{strom84} observations of the central source implied the presence of a pulsar, which was later discovered by \citet{kcb+88} as PSR B1951+32. The pulsar has a dispersion measure (DM) of about 45 cm$^{-3}$~pc~\citep{hlk+04}, and a rotation measure (RM) of $-$182$\pm$8~rad~m$^{-2}$~\citep{wck+04}.

The origin of the morphology of CTB~80 has been puzzling since its identification decades ago. Infrared, H~{\scriptsize I}, and X-ray observations led to the scenario of PSR B1951+32 and its associated pulsar wind nebula (PWN) rejuvenating an old SNR shell~\citep{fesen+1988, koo+93, safi-harb+95}. \citet{castelletti+05} derived a map of spectral indices using observations at 610~MHz and 1380~MHz conducted by \citet{cdg+03}, and found that spectra are flat towards the central PWN area and gradually steepen further away from the PWN area along the shells. This supports that the high-energy electrons accounting for the emission from the shells are supplied by the pulsar and its PWN. 

One of the keys to understand CTB~80 is the magnetic field, which can be studied by radio polarimetry. \citet{mrst85} conducted polarization observations of CTB~80 at 1410~MHz, 1720~MHz, 2695~MHz, and 4750~MHz, and obtained the orientation of magnetic fields projected onto the plane of sky after rotation measure correction. They found that the magnetic field is well ordered, parallel to the eastern and southwestern shell but virtually perpendicular to the northern shell, which motivated them to put forward an alternative scenario of two SNR shells interacting. 

The distance to CTB~80 varies with different ways of measurements. Based on the DM of PSR B1951+32, the distance is either 1.4~kpc~\citep{kcb+88} or 3$\pm$2~kpc~\citep{ymw17} depending on the thermal electron density model. The optical observations yield a color excess $E(B-V)$ of about 0.8, suggesting a distance of about 2.5~kpc~\citep{blair+84}. \citet{szt+18} derived an extinction $A_V$ from the color excess and obtained a distance of 4.6~kpc. From X-ray observations, \citet{mavromatakis+01} determined a column density of H~{\scriptsize I} and then derived a color excess of about 0.4, which they attributed to foreground absorption. This means that the extra color excess of about 0.4 is from the absorption local to the SNR. Assuming a color excess of about 0.4 from the foreground absorption, the distance is about 2~kpc from the recent 3D reddening map by~\citet{green+18}. The H~{\scriptsize I} observations show an absorption feature at the velocity of about ~12~km~s$^{-1}$~\citep{koo+93, lr12}. This velocity is less than the terminal velocity and therefore corresponds to two distances: 1.1 -- 2.1~kpc and 4.0 -- 5.0~kpc after taking into account the velocity dispersion~\citep{lr12}. To be compatible with most of the measurements, we set a value of 2~kpc for the distance to CTB~80 in this paper, which is also applied to PSR B1951+32.

In this paper, we re-examine CTB~80 with more recent polarization observations of higher quality from the Canadian Galactic plane survey at 1420~MHz~\citep{kffu06, lrr+10}, the Effelsberg 2.695-GHz survey of the Galactic plane \citep{drrf99}, and the Sino-German $\lambda$6~cm polarization survey of the Galactic plane \citep{xhr+11}. The layout of the paper is as follows. We present the data in Sect.~2, analysis and discussion on polarization properties in Sect.~3, and conclusions in Sect.~4. 

\section{Data}

\begin{figure}

    \centering
	
	\includegraphics[width=0.48\textwidth]{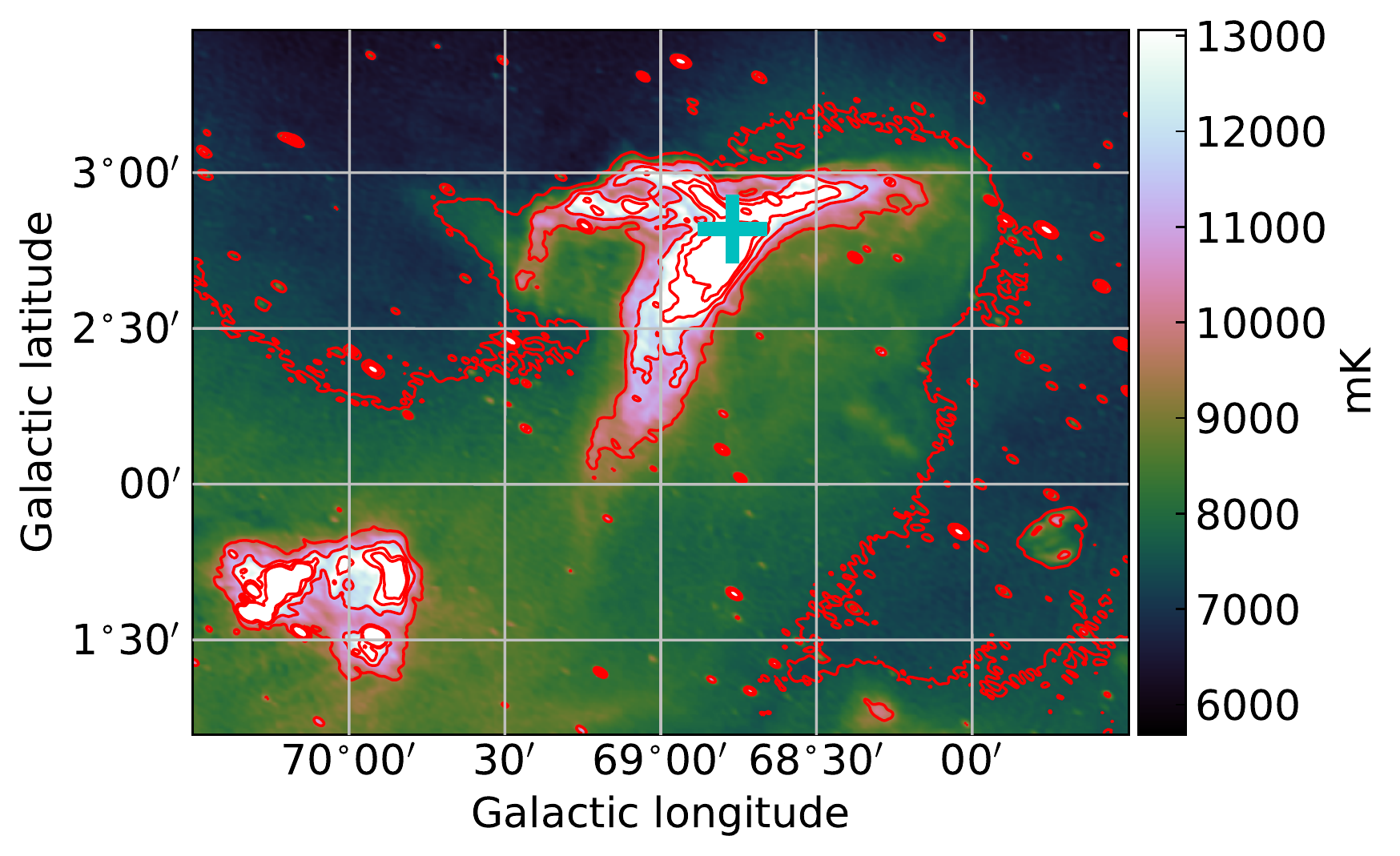}
	\includegraphics[width=0.48\textwidth]{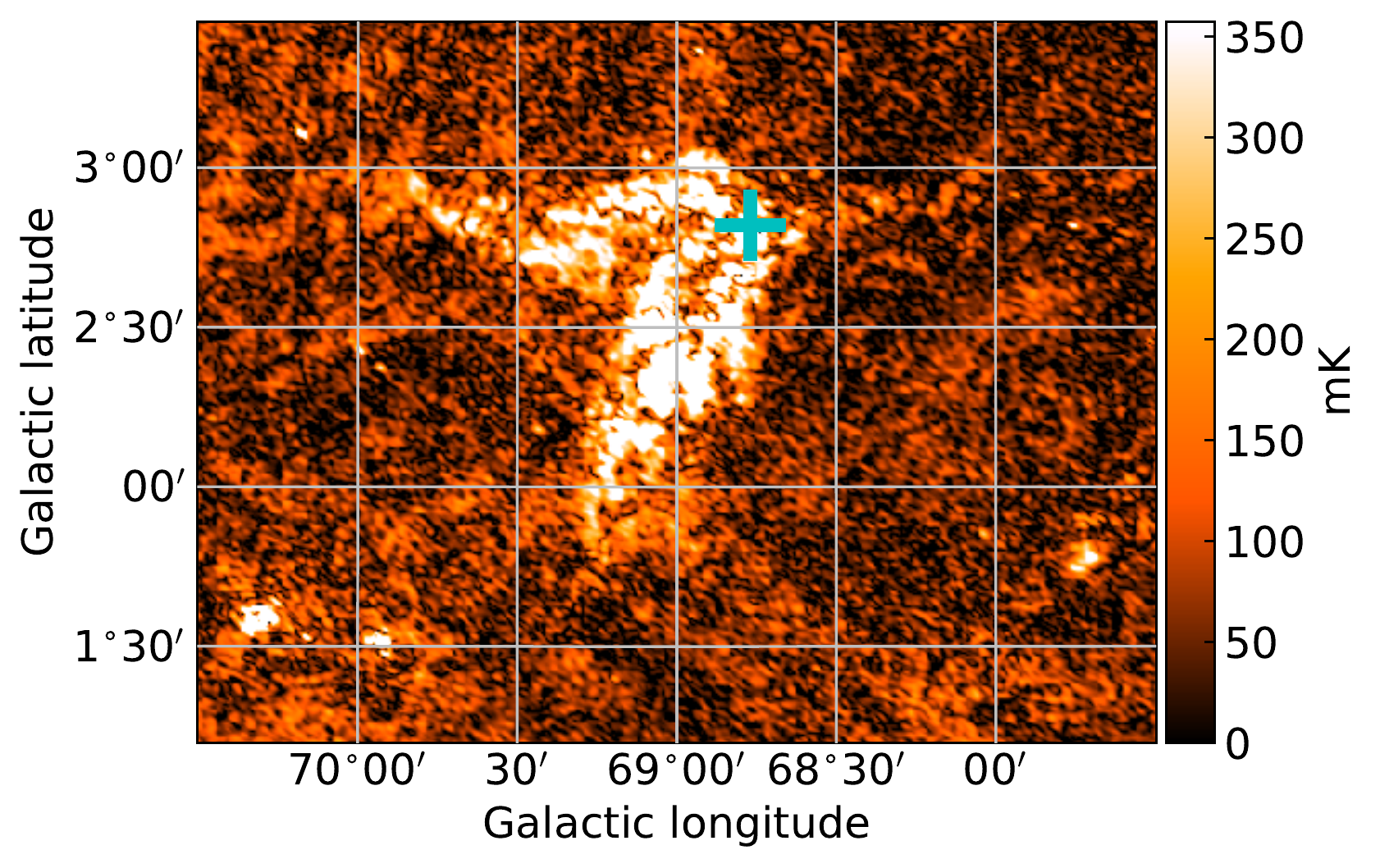}
	\includegraphics[width=0.48\textwidth]{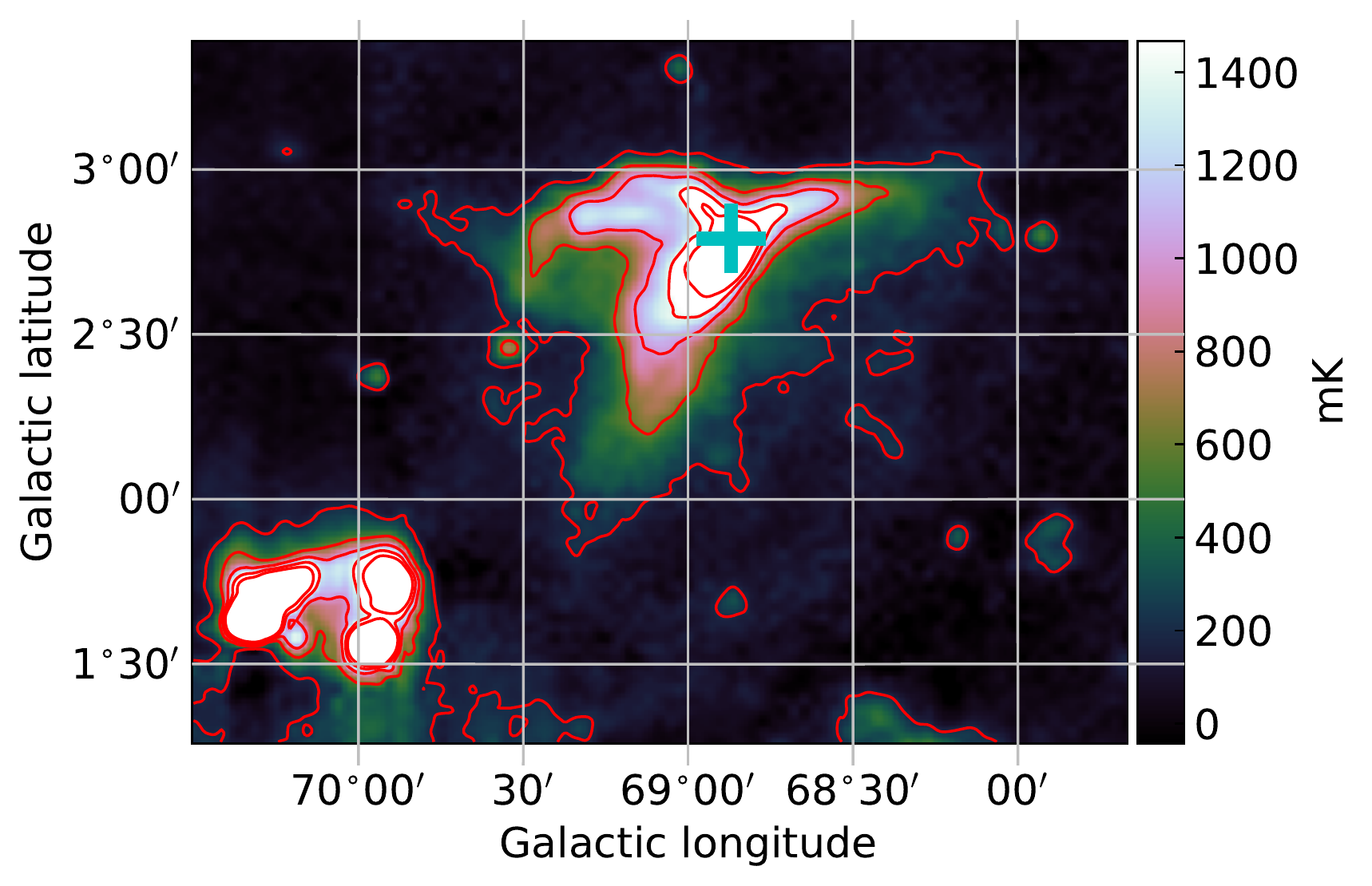}
	\includegraphics[width=0.48\textwidth]{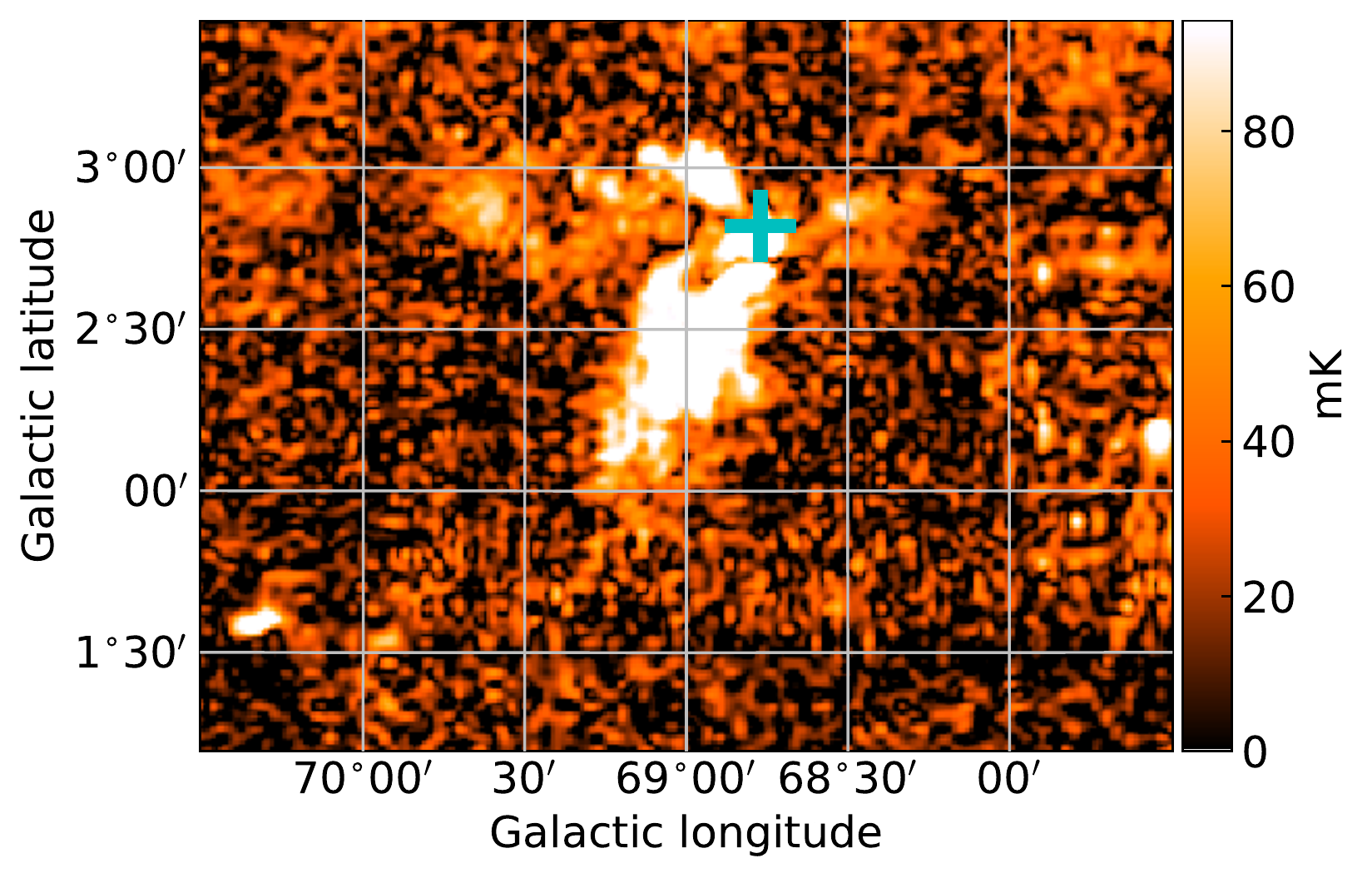}
	\includegraphics[width=0.48\textwidth]{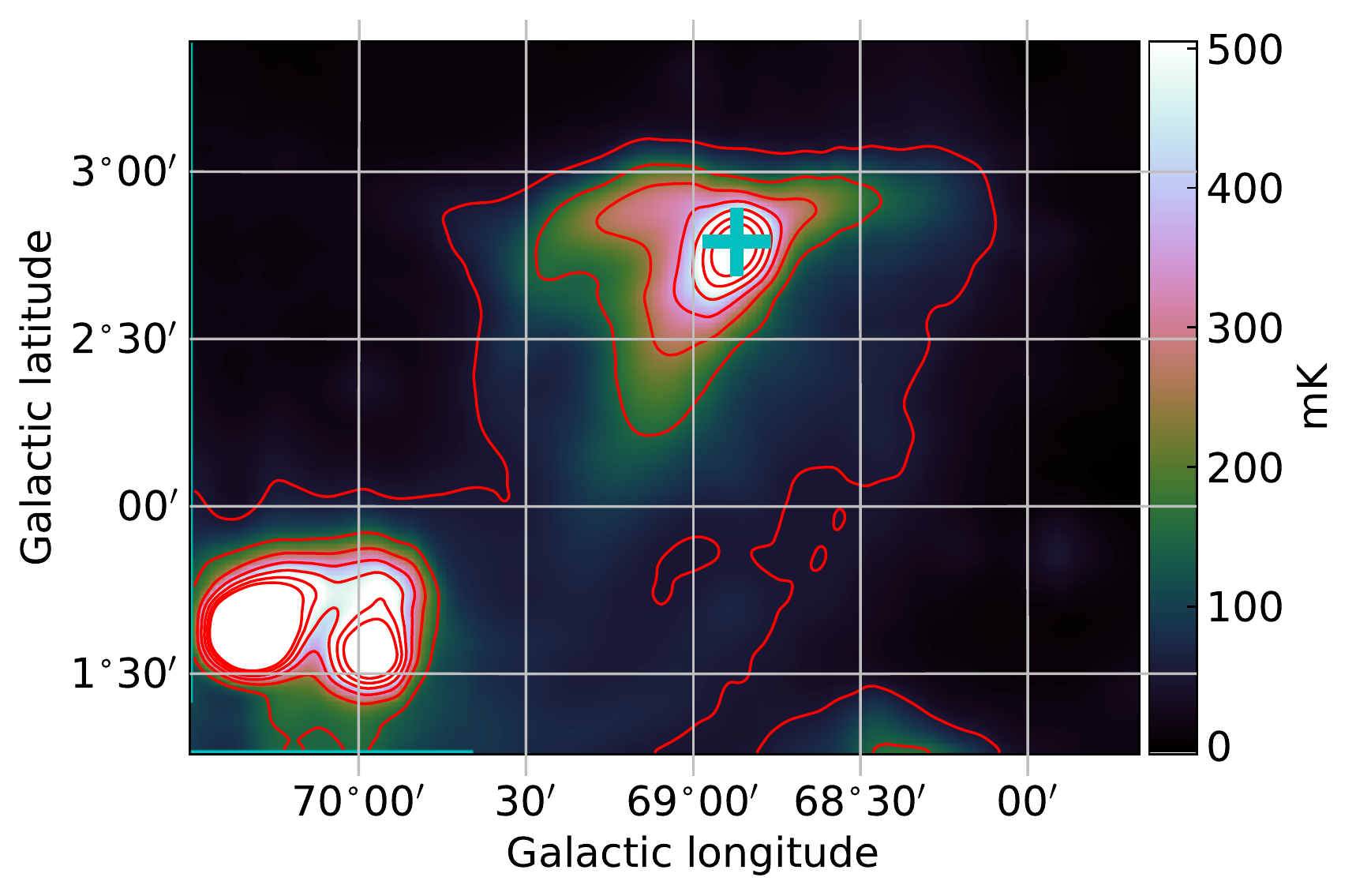}
	\includegraphics[width=0.48\textwidth]{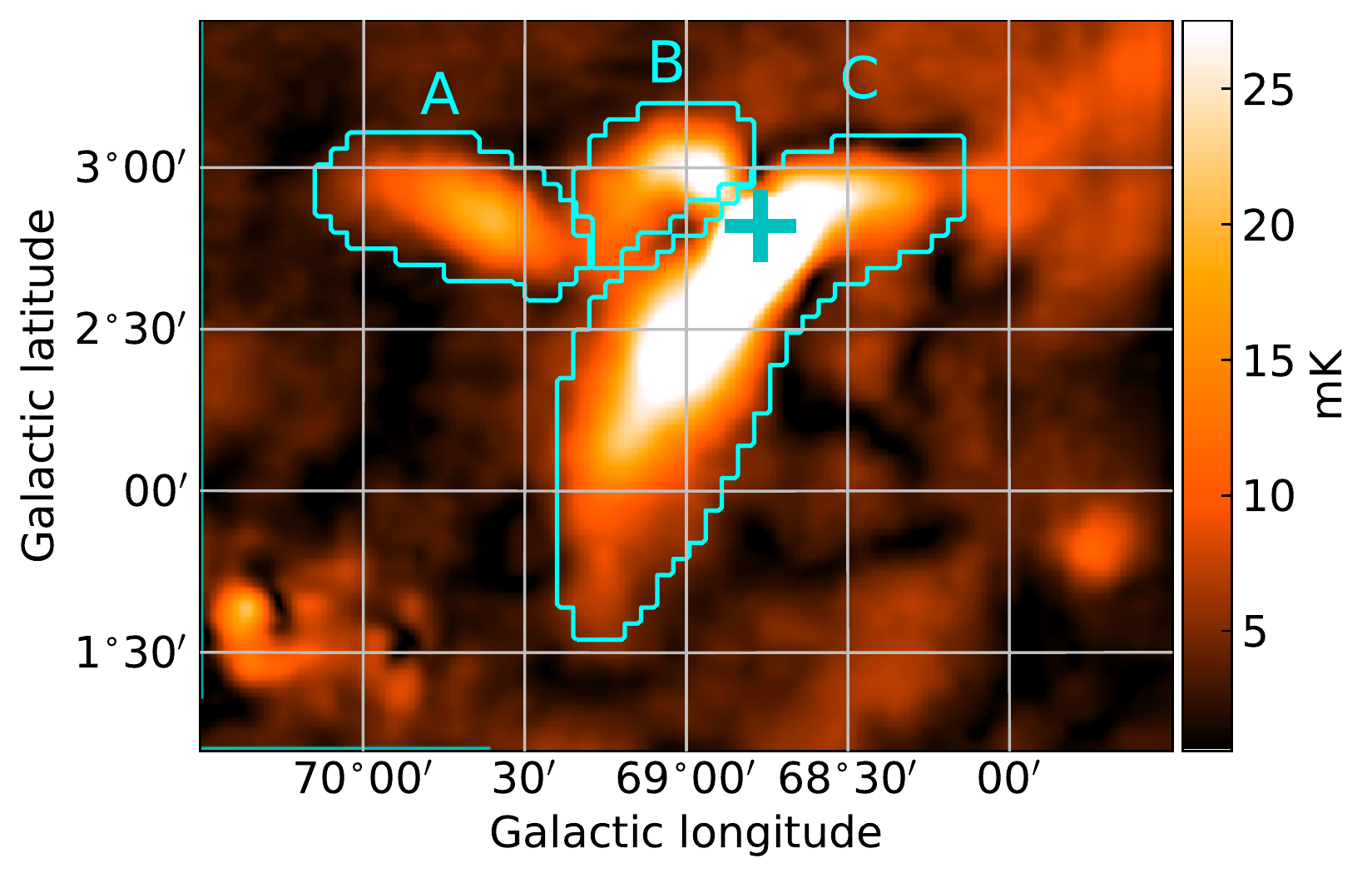}
    \caption{Total-intensity images (left column) and polarized-intensity images (right column) of CTB~80 at 1420~MHz (top) , 2695~MHz (middle), and 4800~MHz (bottom) at angular resolutions as listed in Table~\ref{tab1}. The contour levels start from 7500~mK and increase by 2000~mK at 1420~MHz; start from 200~mK and increase by 400~mK at 2695~MHz; and start from 50~mK and increase by 100~mK at 4800~MHz. All the intensity values are in main beam brightness temperature. The pulsar PSR B1951+32 is marked by a cross in each panel.}
    \label{fig:all_images}
\end{figure}

We focus our analysis on the polarized intensity ($PI$) images of CTB~80 at 1420~MHz, 2695~MHz, and 4800~MHz. We also investigate the total intensity ($I$) images at these frequencies as well as that at 408~MHz for the spectrum. All the data are cutouts from published Galactic plane surveys with characteristics listed in Table~\ref{tab1}. 

\begin{table}
\bc
\begin{minipage}[]{100mm}
\caption[]{Characteristics of the surveys.\label{tab1}}\end{minipage}
\setlength{\tabcolsep}{10pt}
\small
 \begin{tabular}{lllll}
  \hline\noalign{\smallskip}
           & 408   & 1420   &  2695  & 4800  \\
Frequency  & (MHz) & (MHz)  &  (MHz) & (MHz) \\
  \hline\noalign{\smallskip}
Survey              & CGPS & CGPS      & Effelsberg survey   & Sino-German survey\\
Telescope           & DRAO & DRAO      & Effelsberg          & Urumqi\\
Angular resolution  & $3\farcm4\times3\farcm4 \,{\rm cosec}\delta$ & $58\arcsec\times58\arcsec {\rm cosec}\delta$       &  $4\farcm3$     &  $9\farcm5$     \\ 
rms noise, $I$ (mK~$T_{\rm B}$) & 1000 &50      & 20  & 1  \\
rms noise, $PI$ (mK~$T_{\rm B}$) & & 30     & 13  & 0.6 \\
  \noalign{\smallskip}\hline
\end{tabular}
\ec
\tablecomments{0.95\textwidth}{The quoted rms-noise values were measured from the area surrounding CTB~80 and are in main beam brightness temperature mK~$T_{\rm B}$.}
\end{table}

At 408~MHz and 1420~MHz, CTB~80 was covered by the Canadian Galactic Plane Survey (CGPS) with the synthesis telescope at the Dominion Radio Astrophysical Observatory~\citep{taylor+03, lrr+10, Tung+17}. Images of most of the known SNRs from the survey were presented by \citet{kffu06}. The resolution is about 1$\arcmin$. The data can be accessed from the Canadian Astronomy Data Center\footnote{https://www.cadc-ccda.hia-iha.nrc-cnrc.gc.ca/en/cgps/}. The interferometric CGPS data miss large-scale structures and were combined with single-dish data from the Effelsberg 100-m and the Stockert 25-m telescopes at 1420~MHz \citep{rei82, Reich04}. The maps at 2695~MHz were from the Galactic plane survey conducted by using the Effelsberg 100-m telescope in total intensity~\citep{reich9011} and in polarization~\citep{drrf99}, which can be retrieved from the ``Survey Sampler" hosted by the Max-Plank-Institut f\"ur Radioastronomie\footnote{https://www.mpifr-bonn.mpg.de/survey.html}. The original resolution is $4\farcm3$. The data at 4800~MHz were from the Sino-German $\lambda$6~cm polarization survey of the Galactic plane~\citep{xhr+11}. Maps of large SNRs including CTB~80 from this survey were presented by \citet{ghr+11}. The resolution is about $9\farcm5$. The data are also publically available\footnote{http://zmtt.bao.ac.cn/6cm/index.html}. 

The original images of both $I$ and $PI$ for CTB~80 are shown in Fig.~\ref{fig:all_images}. The polarized intensity is calculated from Stokes $Q$ and $U$ as $PI=\sqrt{Q^2+U^2-\sigma^2}$~\citep{wk74}, where $\sigma$ is the rms noise in $Q$ and $U$. To highlight the weak diffuse emission, contours are overlaid on the  total intensity images. From the total-intensity maps, we can clearly see two bright shell structures, where the pulsar and its PWN are located at the intersection of the two shells. The emission from the PWN is blended with that of the shells. The bright complex at the lower left of the maps contains many individual H~{\scriptsize II} regions which are unrelated to CTB~80 \citep{anderson+18}. From the polarization images, two shell structures corresponding to those in the total-intensity images are clearly visible at all three frequencies, and are marked as B and C, respectively. Another shell-like structure extends from shell B, as can be seen at 1420~MHz and 4800~MHz. This partial shell structure is marked as A, which manifests as a diffuse patch at 2695~MHz. The total intensity corresponding to shell A is weak at all three frequencies. 

\section{Results and discussions}

We smoothed the maps of $I$, $Q$, and $U$ at all the frequencies to a common resolution of 10$\arcmin$ for the analyses in this section. 

\subsection{Total intensity spectrum}

We checked the total-intensity spectra by using the TT-plot method~\citep{tpkp62}, namely plotting and linearly fitting brightness temperature at one frequency against that at the other frequency to obtain the spectral index $\beta$ as $T\propto\nu^\beta$. The spectral index $\alpha$ for flux density $S$, defined as $S\propto\nu^\alpha$, can be obtained as $\alpha=\beta+2$. TT-plots are immune to the influence of constant large-scale background emission, which is critical for weak structures such as shell A. From Fig.~\ref{fig:all_images}, it can be clearly seen that CTB~80 sits on a plateau of diffuse emission whose influence on the total-intensity spectra can also be eliminated by using TT-plots. 

\begin{figure}
    \centering
    \includegraphics[width=0.9\textwidth]{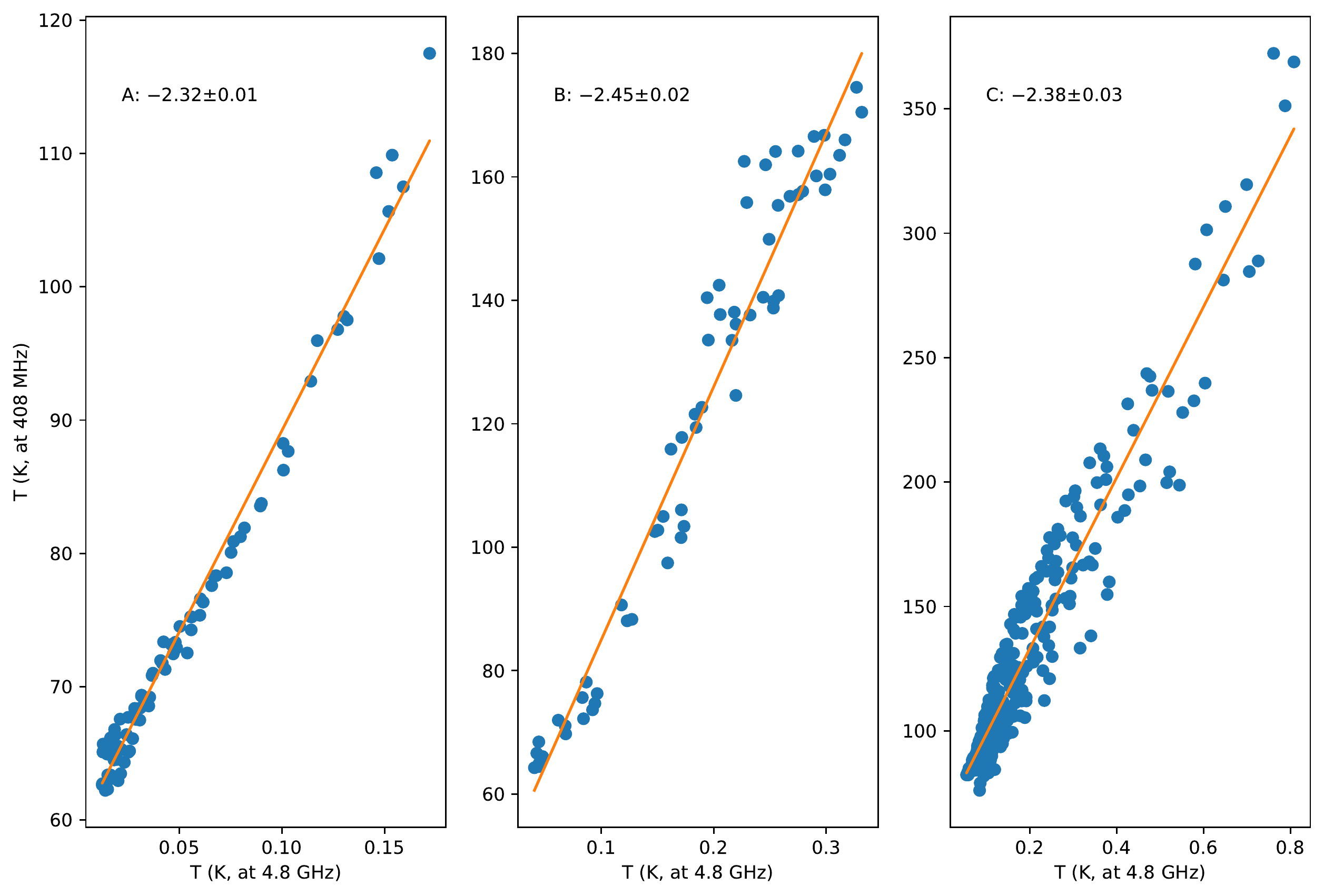}
    \includegraphics[width=0.9\textwidth]{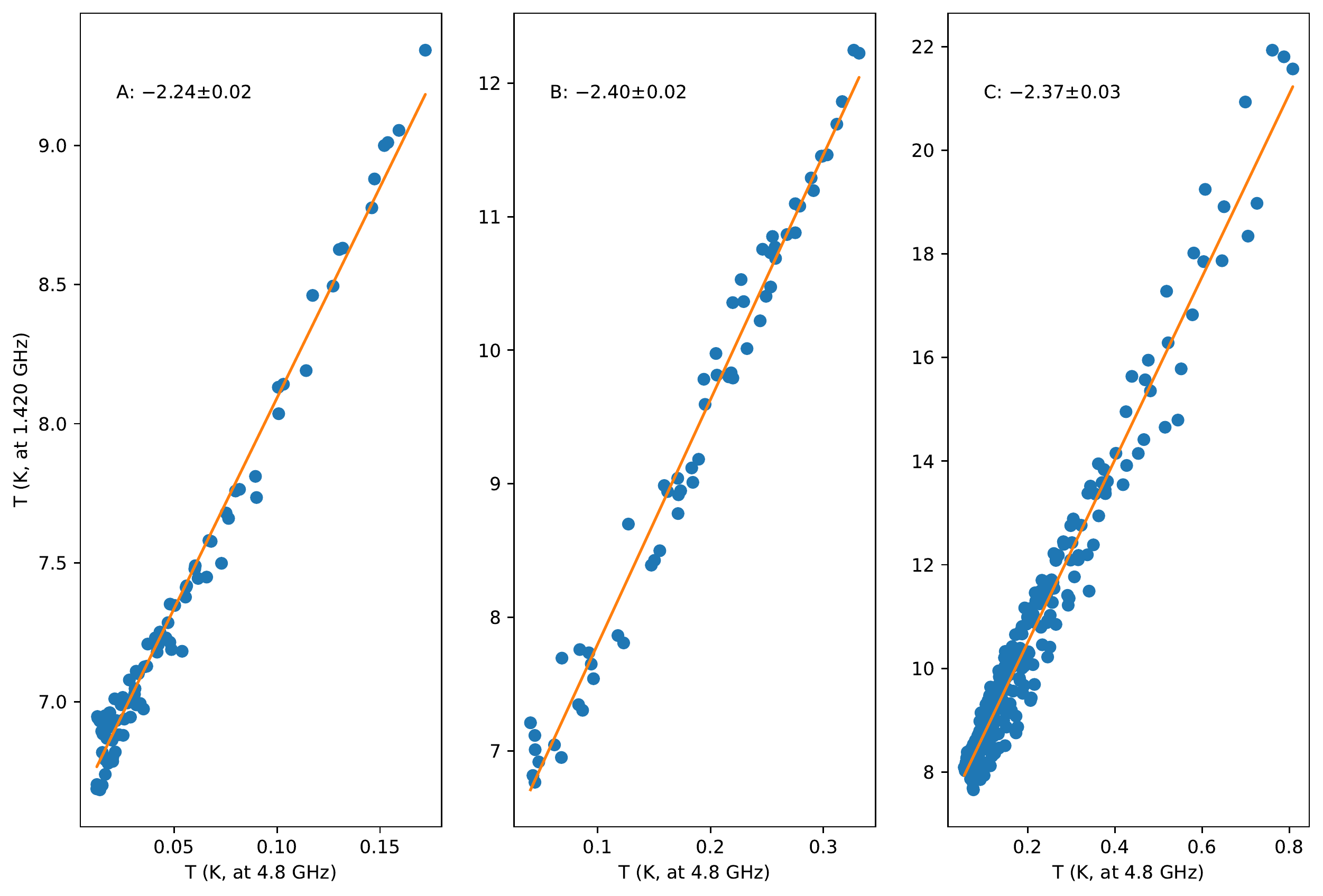}
    \caption{TT-plots for total intensities between 408~MHz and 4800~MHz, and between 1420~MHz and 4800~MHz for the structures A, B, and C.}
    \label{fig:tt}
\end{figure}

The TT-plot results between 408~MHz and 4800~MHz, and between 1420~MHz and 4800~MHz are shown in Fig.~\ref{fig:tt} for all the three areas. Both shell B and shell C have a brightness temperature spectral index of $\beta \sim -2.4$ corresponding to a flux density spectral index of $\alpha \sim -0.4$, consistent with previous results~\citep{kffu06, ghr+11}. For the shell structure A, the spectrum is flatter with a brightness temperature spectral index of $\beta \sim -2.3$ between 408~MHz and 4800~MHz, and $\beta \sim -2.2$ between 1420~MHz and 4800~MHz. 

\subsection{Depolarization}\label{sect:dp}

\begin{figure}
    \centering
    \includegraphics[width=0.49\textwidth]{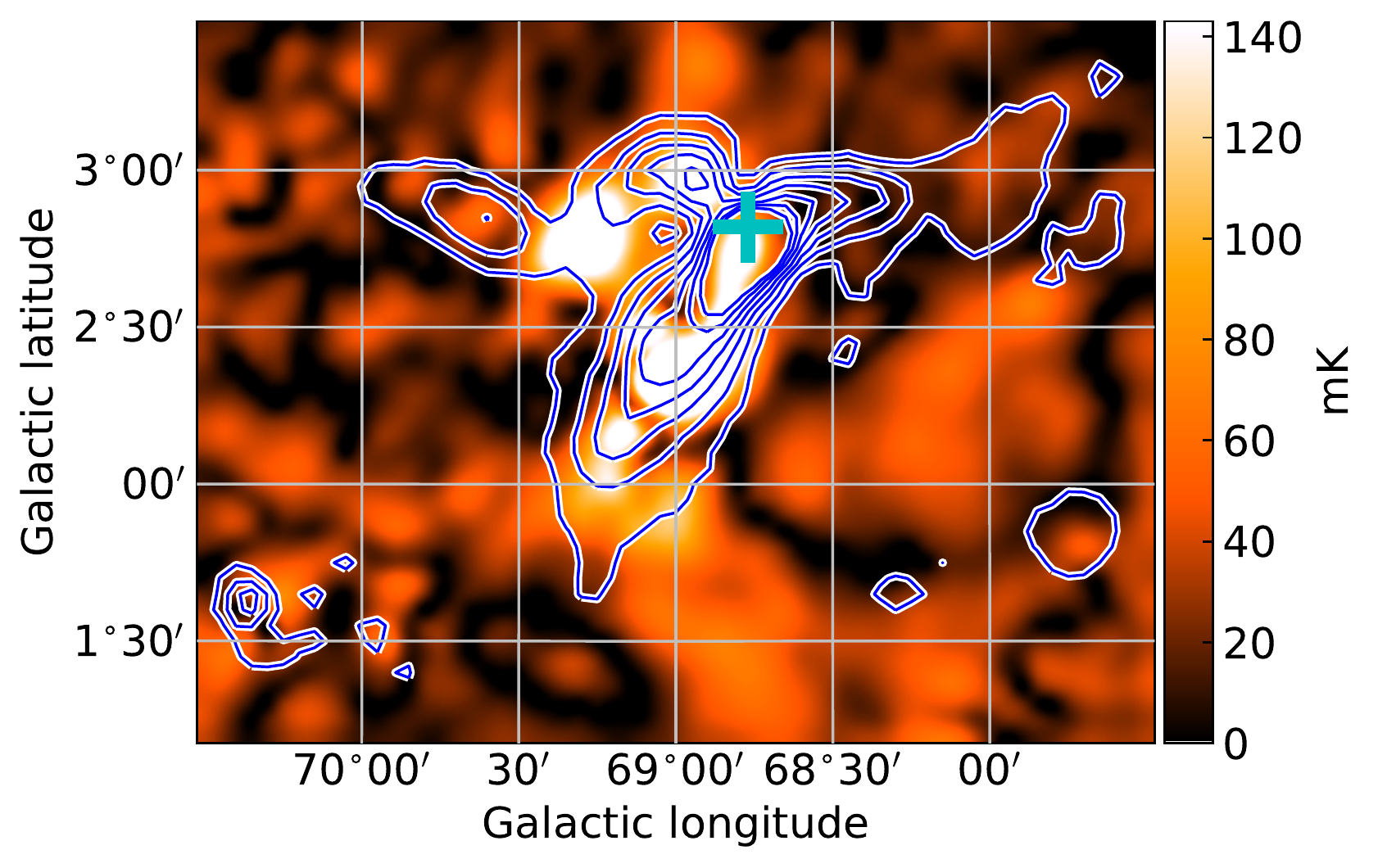}
	\includegraphics[width=0.49\textwidth]{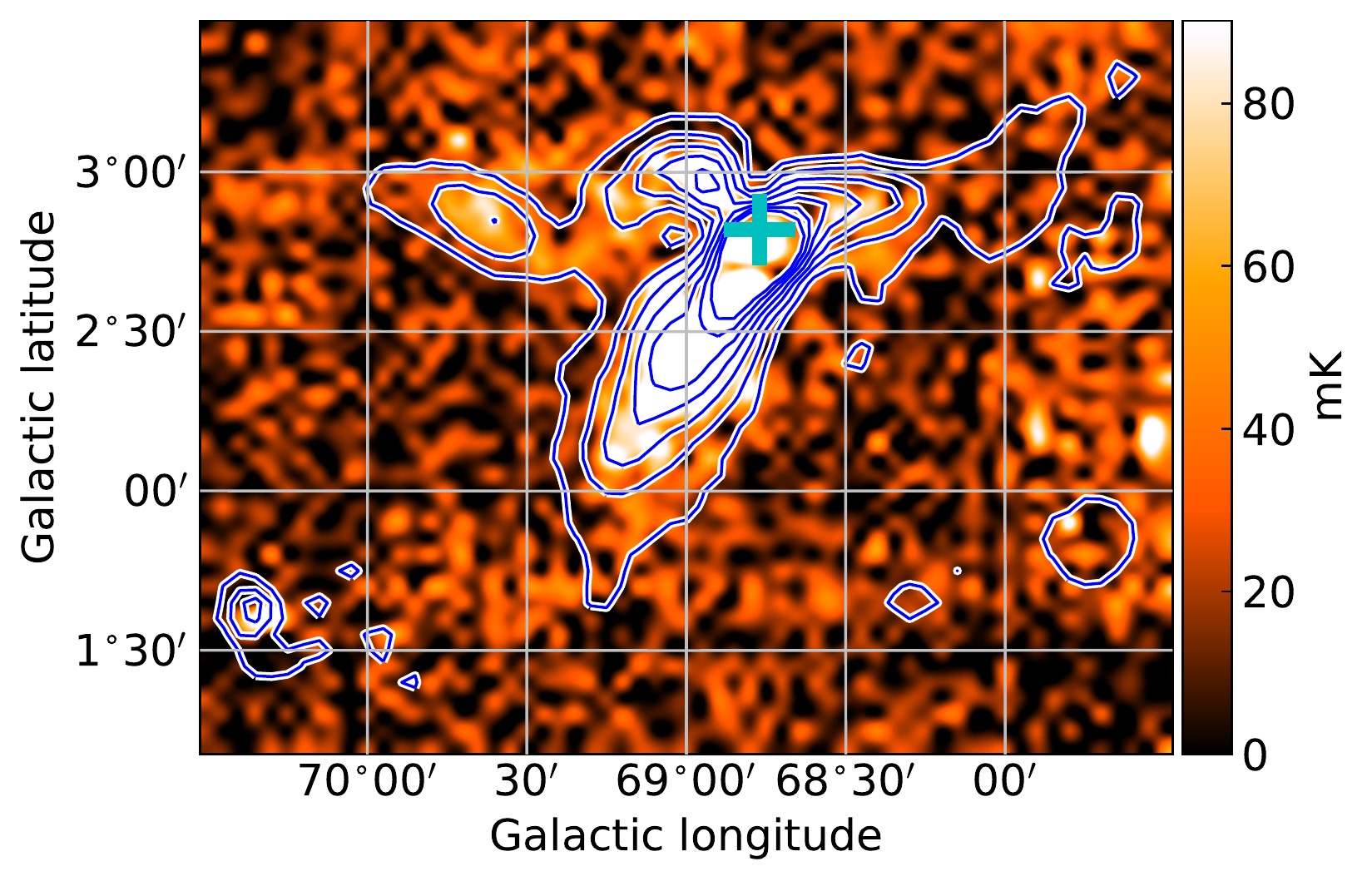}
	\includegraphics[width=0.48\textwidth]{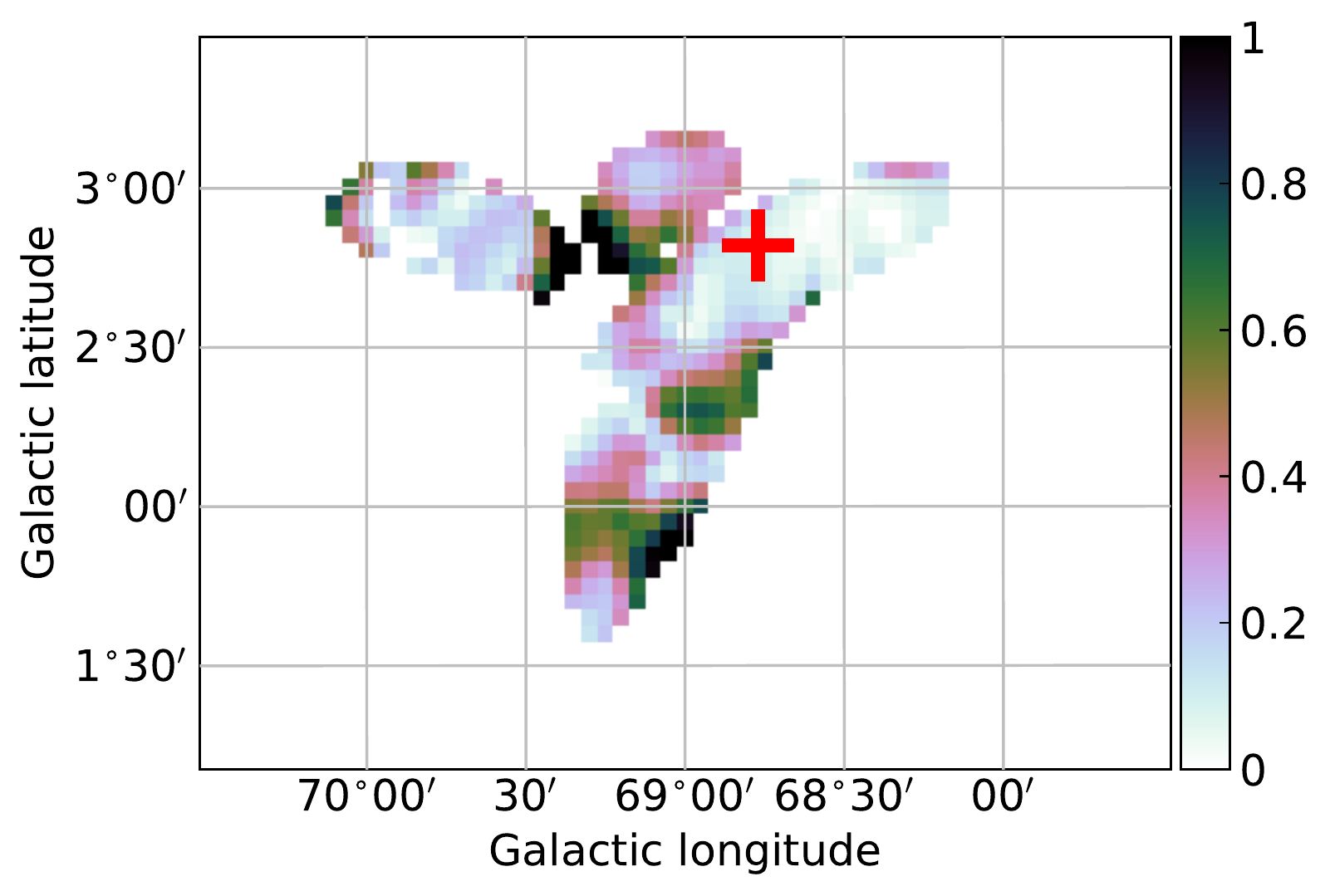}
	\includegraphics[width=0.48\textwidth]{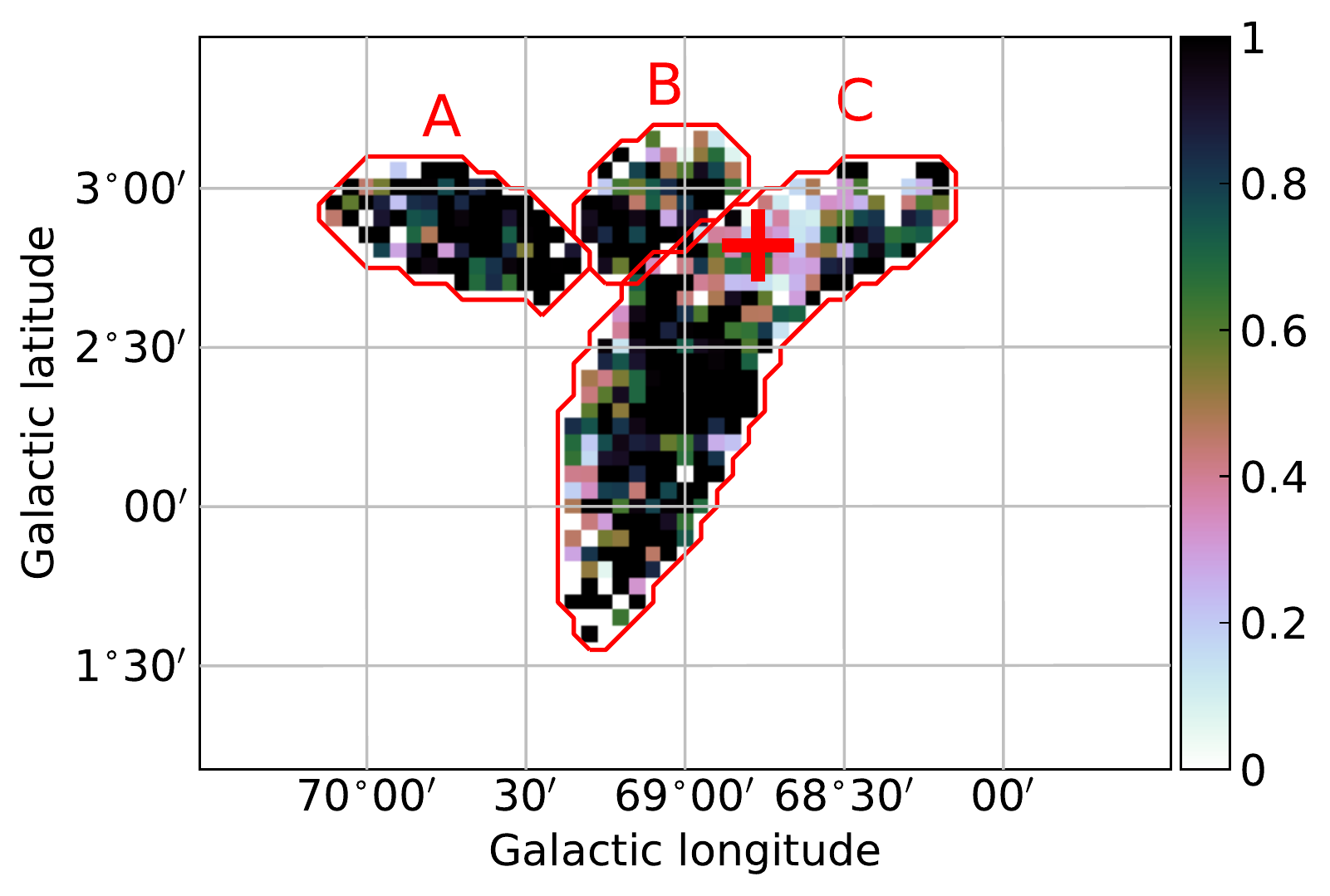}
	    \caption{Maps of the polarized intensity of CTB~80 at 1420~MHz (top left panel) and at 2695~MHz (top right panel) 
	    at $10\arcmin$ angular resolution, and the distribution of the depolarization factor (see text) at 1420~MHz (bottom left panel) and at 
	    2695~MHz (bottom right panel). The contours indicating polarized intensity at 4800~MHz runnning from 6~mK to 42~mK in steps of 6~mK. The pulsar 
	    PSR B1951+32 is marked by a cross in all panels.}
    \label{fig:dp}
\end{figure}

The emission encoded in the maps in Fig.~\ref{fig:all_images} consists of contributions from CTB~80 and the Milky Way background. The latter is irrelevant and shall be removed for further analysis. For $Q$ and $U$ maps, we removed the background emission by subtracting a plane fitted with the values surrounding CTB~80. From these reprocessed $Q$ and $U$ maps we obtained polarized intensity $PI$~(Fig.~\ref{fig:dp}), and polarization angle as $\psi=\frac{1}{2}{\rm arctan}\frac{U}{Q}$. 

For the total-intensity maps, it is very difficult to exclude the background emission. As a consequence, the polarization percentage defined as $PC=PI/I$ would have large uncertainties. In order to circumvent the problem, we compare the polarization at different frequencies by using the relative polarization percentage or the depolarization factor defined as $DP_\nu=\frac{PI_\nu}{PI_{\nu_0}}(\frac{\nu_0}{\nu})^\beta$, where $\nu=1420$~MHz or 2695~MHz, and $\nu_0=4800$~MHz. Here we assume that the depolarization is the least at 4800~MHz and use $PI_{4800}$ as the reference. If there is no depolarization, both $DP_{1420}$ and $DP_{2695}$ would be around 1.  

We calculated $DP_{1420}$ and $DP_{2695}$ by using the brightness temperature spectral index from the TT-plots between 1420~MHz and 4800~MHz~(Fig.~\ref{fig:tt}), namely $\beta=-2.24$ for the area A, $\beta=-2.40$ for the area B and $\beta=-2.37$ for the area C. The results are shown in the lower panels of Fig.~\ref{fig:dp}. 

There is strong depolarization at 1420~MHz. The general morphology of the polarized intensity at 1420~MHz before smoothing resembles that at 4800~MHz~(Fig.~\ref{fig:all_images}). However, the polarized intensity is largely reduced and the morphology dramatically changes after smoothing because of the fluctuation of the polarization angles. There are clear offsets between positions of strong polarized emission at 1420~MHz and that at 4800~MHz, as can be seen from Fig.~\ref{fig:dp}. This implies that the polarized emission at 1420~MHz and at 4800~MHz probably originate from different regions. We therefore did not include the polarization data at 1420~MHz for RM calculations.

The polarized intensity at 2695~MHz corresponds well to that at 4800~MHz. Thus, the depolarization factor is around 1 for all areas except for the region near PSR B1951+32 where $DP_{2695}$ varies from about 0.1 to about 0.7, indicating depolarization towards this region. A detailed discussion on the depolarization mechanisms was provided by \citet{sbs+98}. If the depolarization is caused by the thermal gas that co-exists with the synchrotron-emitting medium, which is referred to as depth depolarization, the depolarization factor can be expressed as $DP_\nu=\frac{\sin({\rm RM}\lambda^2)}{\sin({\rm RM}\lambda_0^2)}\frac{\lambda_0^2}{\lambda^2}$. Here the wavelengths $\lambda$ and $\lambda_0$ corresponds to the frequencies 2695~MHz and 4800~MHz, respectively; and RM is the rotation measure contributed by the whole medium. Given the $DP_{2695}$ values above for the area near PSR B1951+32, the absolute values of RM fall in the range of $\sim$120 -- $\sim$240~rad~m$^{-2}$. If the depolarization is caused by RM fluctuations $\sigma_{\rm RM}$ within the beam, which is referred to as beam depolarization, the depolarization factor can be written as $DP_\nu=\exp[-2\sigma_{\rm RM}^2(\lambda^4-\lambda_0^4)]$. The values of $DP_{2695}$ yield the RM fluctuations from $\sim$35~rad~m$^{-2}$ to $\sim$85~rad~m$^{-2}$. However, the RM fluctuations at these levels will cause virtually complete depolarization at 1420~MHz even before smoothing, which contradicts the observations. Therefore, we attribute the observed depolarization around the pulsar area to depth depolarization.

\subsection{RM map}

The polarization angle varies with wavelength $\lambda$ because of Faraday rotation, and can be expressed as $\psi=\psi_0+{\rm RM}~\lambda^2$. RM is equal to the Faraday depth and thus is calculated as ${\rm RM}=0.81\int_{\rm source}^{\rm observer}n_e B_\parallel {\rm d}l$, where $n_e$ is the thermal electron density in cm$^{-3}$, $B_\parallel$ is the line-of-sight component of the magnetic field in $\mu G$, ${\rm d}l$ is the path length segment along the line of sight in pc, and the resultant RM is in rad~m$^{-2}$. Positive RMs mean that the magnetic field points towards the observer. The intrinsic polarization angle $\psi_0$ is perpendicular to the orientation of the transverse magnetic fields in the source.

\begin{figure}
    \centering
    \includegraphics[width=0.9\textwidth]{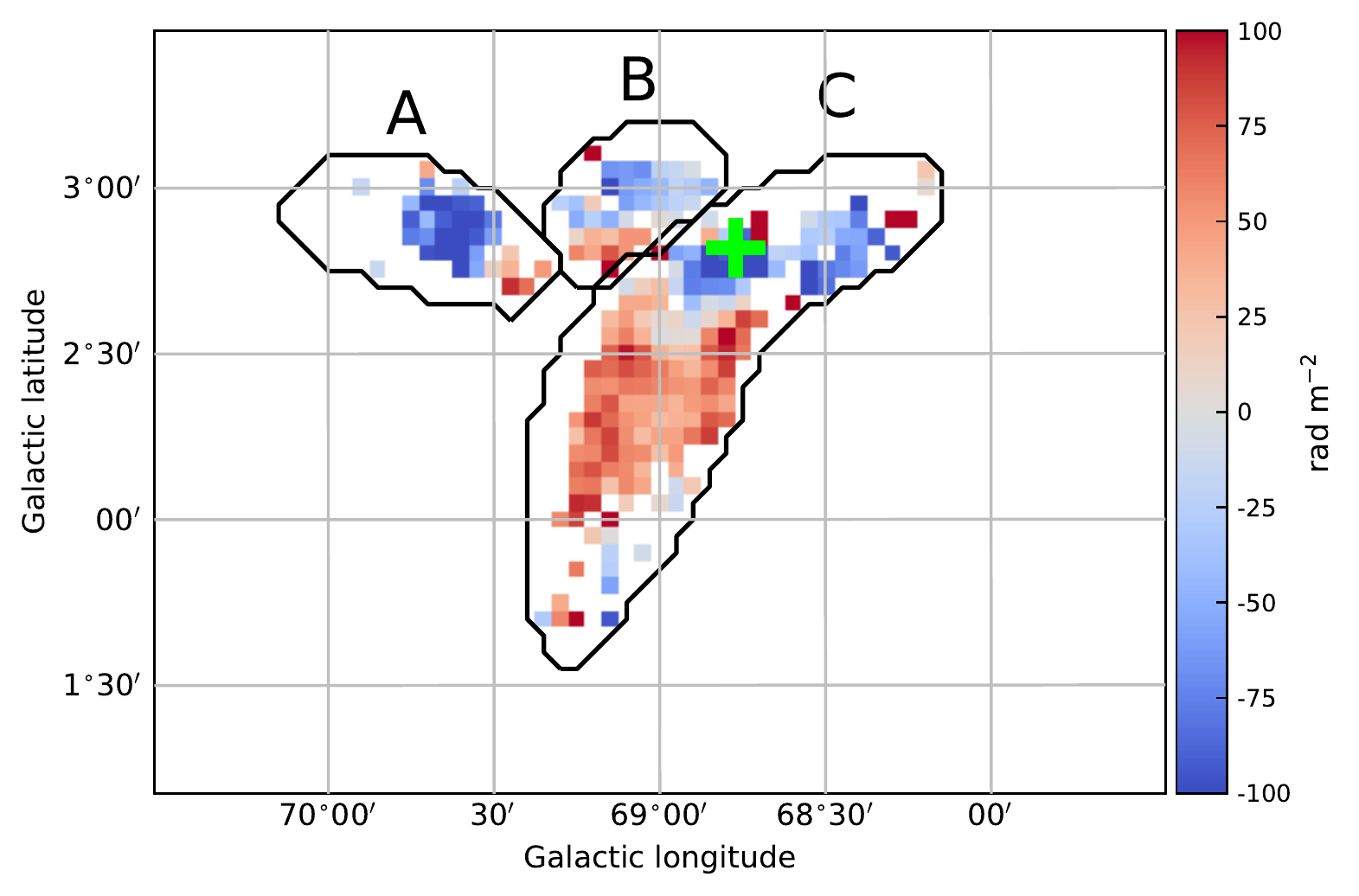}
    \includegraphics[width=0.9\textwidth]{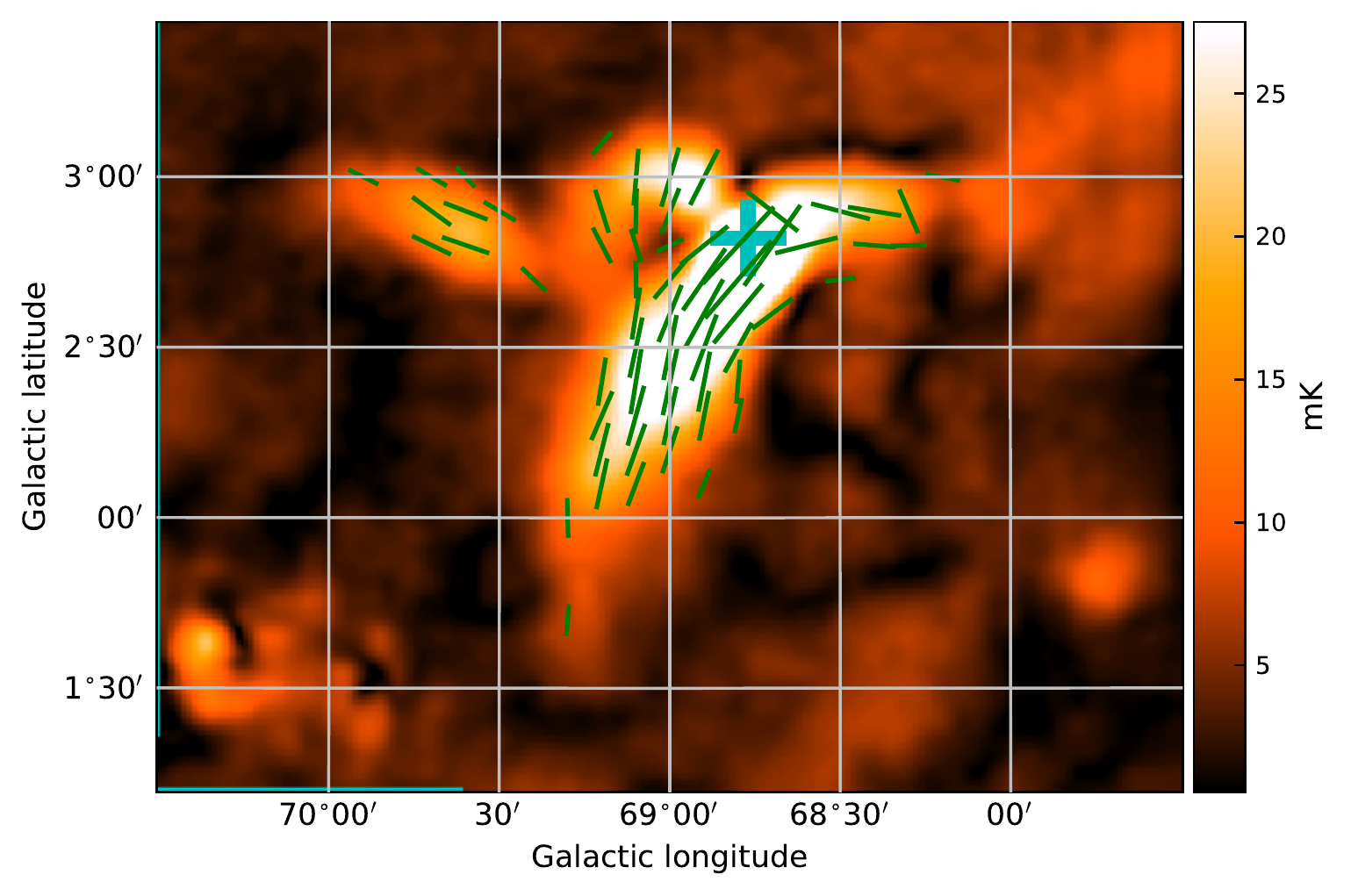}
    \caption{RM (top panel) and orientation of transverse magnetic fields indicated by bars overlaid on the polarized-intensity image at 4800~MHz (bottom panel). The length of the bars is proportional to the square root of polarized intensity. The crosses mark the position of PSR B1951+32.}
    \label{fig:rm_B}
\end{figure}

We used data at 4800~MHz and 2695~MHz to estimate the rotation measure as ${\rm RM}=\frac{\psi_2-\psi_1}{\lambda_2^2-\lambda_1^2}$, where the subscripts 1 and 2 stand for quantities at these two wavelengths, respectively. Because of the $n\pi$ ambiguity of polarization angles, the resultant RM can differ by an integer multiple of $\mathcal{R}=\pi/(\lambda_1^2-\lambda_2^2)\sim370$~rad~m$^{-2}$. We assume RMs with the minimal absolute values as relevant. RMs calculated that way are shown in Fig.~\ref{fig:rm_B} (top panel). All pixels with polarized intensity smaller than $3\times\sigma$-level, which is about 36~mK at 2695~MHz and 1.5~mK at 4800~MHz, were not included for the RM calculation. The intrinsic polarization angles $\psi_0$ were then determined based on RM. The orientation of magnetic fields perpendicular to the line of sight, or transverse magnetic fields,  is also shown in Fig.~\ref{fig:rm_B}. The transverse magnetic fields nearly follow the shell C and are almost perpendicular to the shell B, which is consistent with the results by \citet{mrst85}. 

The RM map exhibits clear patterns as can be seen from Fig.~\ref{fig:rm_B}. For the shell structure C, RM changes from positive values towards the southeast to negative values towards the northwest, where the transition is around Galactic latitude $b=2.7\degree$, which confirms the results by \citet{mrst85}. The RM 
distribution is fairly smooth for the positive RM-area with an average of about +60~rad~m$^{-2}$. For the higher latitude area, the part encompassing the pulsar 
and its PWN has an average RM of about $-$150~rad~m$^{-2}$, which is close to the RM of the pulsar PSR B1951+32. The remaining part further northwest has an 
average RM of about $-$60~rad~m$^{-2}$.  

The RM distribution towards CTB~80 consists of contributions from the Galactic medium in front of the SNR (RM$_{\rm fg}$) and the medium local to the SNR (RM$_{\rm snr}$), which can be written as $\rm RM = RM_{fg} + \frac{1}{2}RM_{snr}$. Here $\rm RM_{snr}$ represents the integral of the line-of-sight magnetic field weighted by the thermal electron density over the whole SNR, and the factor $\frac{1}{2}$ comes from the assumption that the thermal gas and emitting medium are uniformly mixed~\citep[e.g.][]{sbs+98}. In contrast, the RM of PSR B1951+32 can be expressed as $\rm RM_{psr} = RM_{fg} + {\it f} RM_{snr}$, where the factor $f$ depends on the relative location of the pulsar inside the SNR along the line of sight. If the pulsar sits in the middle of the emitting area, we would expect $f$ to be $\frac{1}{2}$. 

\begin{figure}
    \centering
    \includegraphics[width=0.9\textwidth]{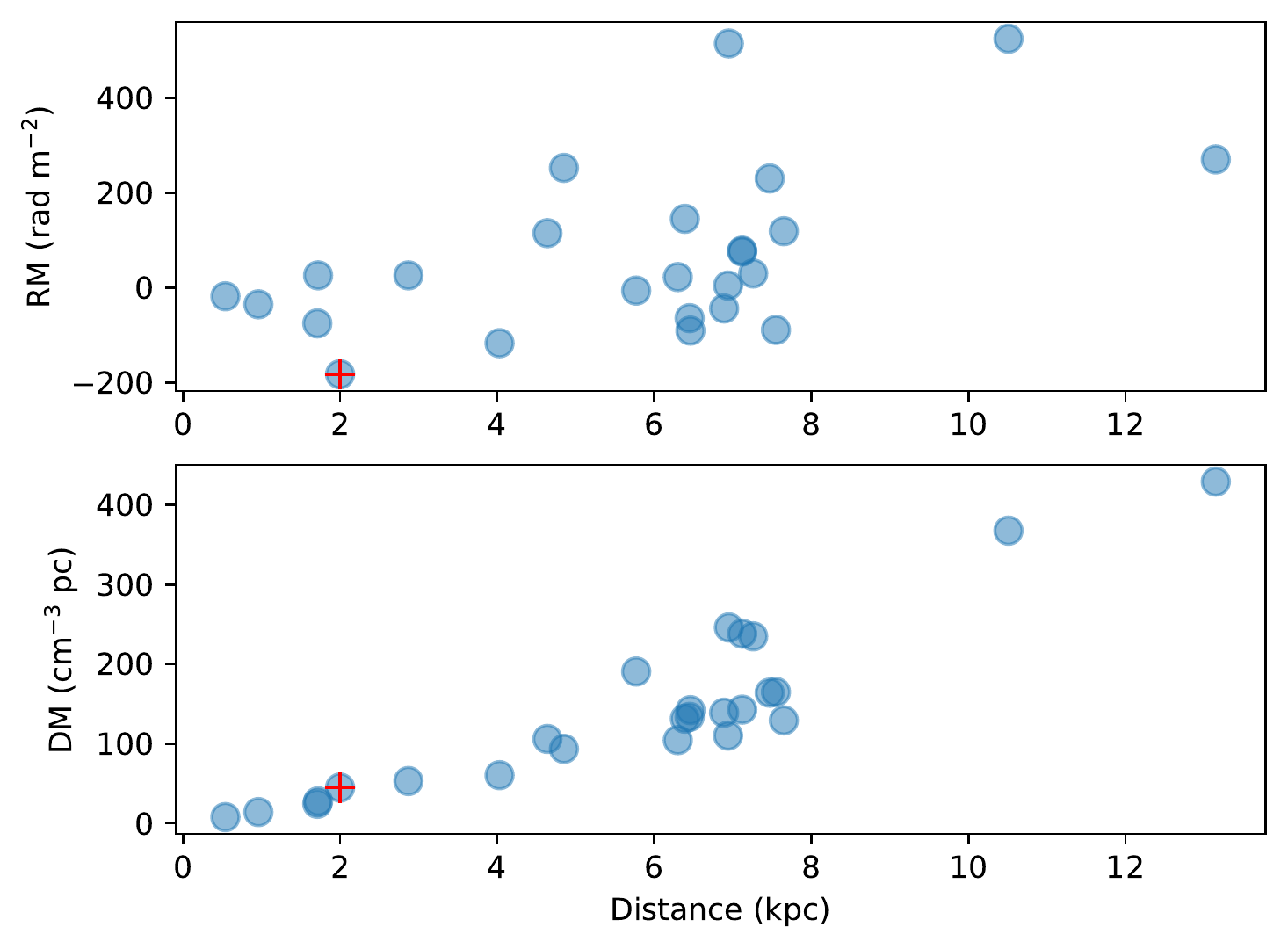}
    \caption{RM and DM versus distance for pulsars within $10\degree$ of PSR B1951+32 which is marked by a cross.}
    \label{fig:psr}
\end{figure}

The foreground RM can be estimated from pulsars and simulations. We retrieved pulsars within $10\degree$ of PSR B1951+32 from the ATNF pulsar catalog\footnote{https://www.atnf.csiro.au/research/pulsar/psrcat/} \citep{mhth05}, and plotted RM and DM against distance in Fig.~\ref{fig:psr}. The distances were calculated from DM according to the electron density model by \citet{ymw17}. For pulsar PSR B1951+32, we used a distance of 2~kpc, the same as for CTB~80. DM is defined as the integral of electron density along the line of sight from the source to the observer. Assuming a uniform distribution of electron density and magnetic field, both DM and RM are expected to have a linear relation with distance. This is roughly the case for DM, which can be seen from Fig.~\ref{fig:psr}. However, RM is complicated with a large scattering. For the four pulsars with distances less than 2~kpc, their RM values are: $-18$~rad~m$^{-2}$, $-35$~rad~m$^{-2}$, $-75$~rad~m$^{-2}$, and $+26$~rad~m$^{-2}$ in the order of increasing distances. The average RM for the four pulsars is about $-26$~rad~m$^{-2}$ with a large standard deviation of about $36$~rad~m$^{-2}$. We run also simulations of 3D Galactic RM towards $(l,\,b)=(69\degree, 2.7\degree)$ with the models of Galactic magnetic fields developed by \citet{sr10}, and obtained a profile of RM versus distance from the average within a radius of $2\degree$. The profile shows that the foreground RM is about $-$40~rad~m$^{-2}$ at a distance of 2~kpc. We below use a value of $-$30~rad~m$^{-2}$ for $\rm RM_{fg}$, which is roughly the average of the estimates from the pulsars and the simulations. It should be emphasized here that $\rm RM_{fg}$ is uncertain though.  

Based on the RM of CTB~80~(Fig.~\ref{fig:rm_B}) and $\rm RM_{fg}$, we obtain a value of about $-240$~rad~m$^{-2}$ for $\rm RM_{snr}$ of the area near PSR B1951+32, falling in the range estimated from depolarization analysis. For PSR B1951+32, we can derive the $f$ value of about $\frac{5}{8}$, slightly larger than $\frac{1}{2}$, meaning that the pulsar is located further than the middle of the emitting medium along the line of sight. For the lower part of shell C, the resulted $\rm RM_{snr}$ is about $+$180~rad~m$^{-2}$. This value is expected to cause large depolarization at 2695~MHz, which is not seen from the depolarization map in Fig~\ref{fig:dp}. Thus synchrotron emission and thermal gas can not be mixed. A Faraday screen with an RM of about $+$90~rad~m$^{-2}$ in front of the synchrotron emission of the lower part of C can explain the observations.

The change of RM signs for SNRs was reported previously for the SNR CTA~1~\citep{srw+11} and SNR G296.5+10.0~\citep{hgk+10}. The former was interpreted by the influence of a foreground cloud in accordance with RMs of extragalactic sources, and the latter was attributed to the toroidal field generated by the stellar wind from the progenitor star. In the case of CTB~80, the foreground RM is small. However, the large positive RMs in the lower part of shell C could be attributed to the thermal gas inside the SNR but in front of the emitting medium, which acts as a Faraday screen. There are no anomalies from RMs of extragalactic sources towards this area when investigating the Galactic foreground RM map by \citet{oppermann+15}. The reason could be that this foreground map was constructed mainly from the RM catalogue by \citet{tss09} with a source density of about 1 per square degree, and is therefore not sensitive to RM variations over smaller scales as in shell C. For CTB~80, the sign change occurs within one shell, and the second scenario thus cannot work. \cite{kb09} demonstrated that the sign of RM changes along the shell of an SNR, in particular when the angle between the magnetic field and the line of sight is small. This seems possible because the location of CTB~80 is close to the strong Cygnus~X region, which results from the tangential view along the local arm with the magnetic field aligned.

\begin{figure}
    \centering
    \includegraphics[width=0.9\textwidth]{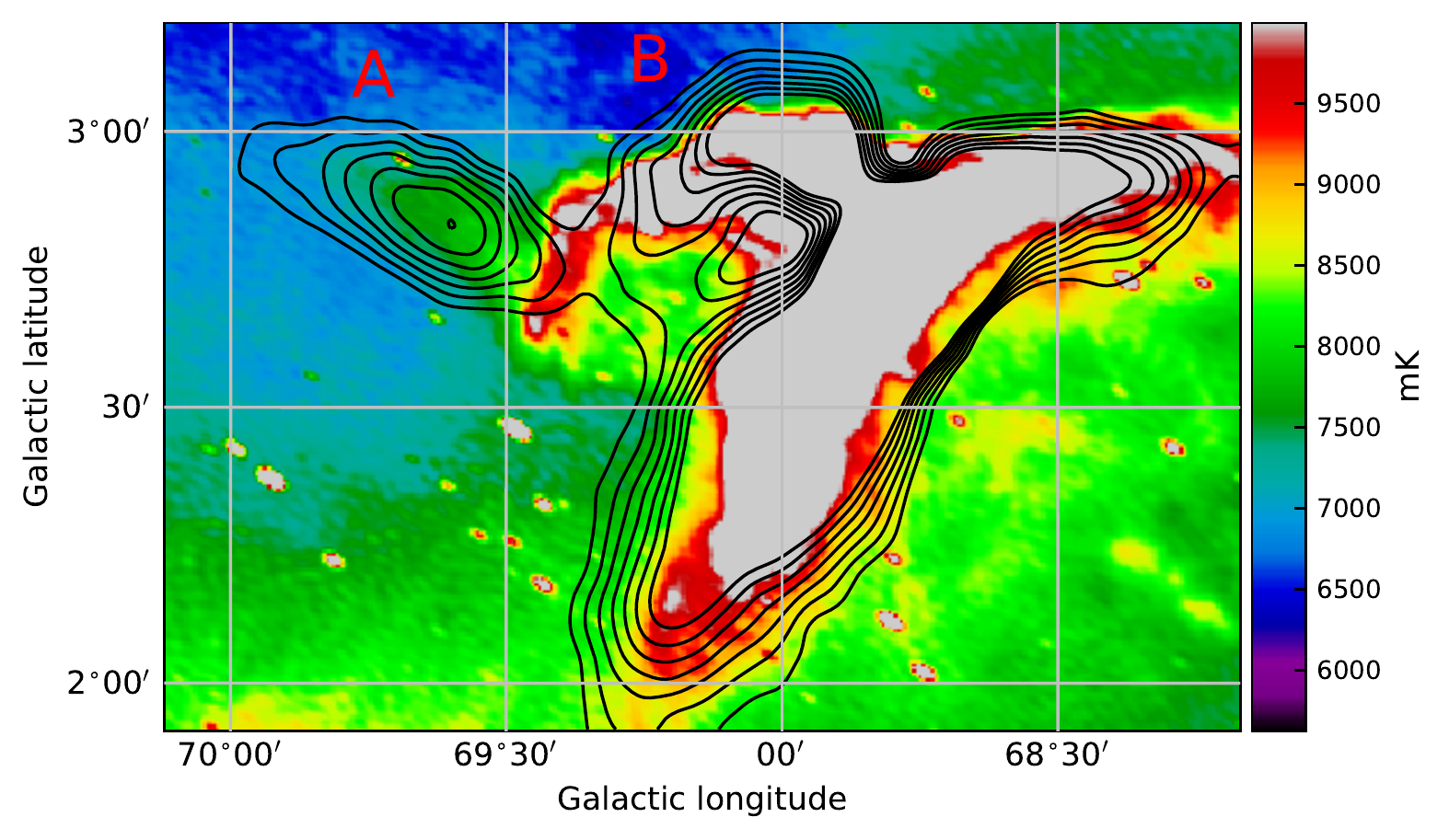}
    \caption{Polarized intensity at 4800~MHz in contours overlaid on the total intensity image at 1420~MHz. The levels of the contours go from 8~mK to 20~mK in steps of 2~mK in main beam brightness temperature.}
    \label{fig:A}
\end{figure}

\subsection{Nature of the structure A}

The partial shell structure A (Fig.~\ref{fig:all_images}) is very intriguing. We overlaid the contours of the polarized intensity at 4800~MHz onto the high resolution total intensity image at 1420~MHz in Fig.~\ref{fig:A}. CTB 80 has an unusual and complex morphology compared to other SNRs and its full extent is not very clear. Whether the faint but highly polarized structure A is part of CTB 80 or not, is difficult to decide. With regard to total intensity, it seems that the shell B bends towards longitudes larger than about $69.4\degr$ to form a quasi-circular shell, whereas shell A is roughly perpendicular to shell B. The non-smooth transition of emission from shell A to shell B suggests that shell A might not be related with shell B and thus CTB~80. The polarized emission from structure A (Fig.~\ref{fig:A}) appears to be separated from CTB~80. However, the end of the total-intensity shell B is seen in superposition with shell A in the unpolarized area that may indicate depolarization. In fact, \citet{mavromatakis+01} found faint emission in H$\alpha$ + [N {\scriptsize II}] and [S {\scriptsize II}] in this area. This means the existence of ionised matter which might cause depolarization through Faraday effects. In the case that depolarization took place, shell A would be located behind shell B and might be connected to the stronger filaments or shells of CTB~80.

Both shell B and shell C are partly fueled by receiving high-energy particles from PSR B1951+32 and its PWN~\citep[e.g.][]{castelletti+05}. Since structure A is further apart, we would expect its spectrum to be similar or steeper than that of shell B or C. This contradicts our result where structure A has a flatter spectrum. The flattening of a spectrum could be caused by strong compression by interaction with the interstellar medium. The flux-density spectral index can be connected to the shock compression ratio $r$ following $\alpha=-3/2(r-1)$~\citep{reynolds+12}. Given $\alpha=-0.2$ for structure A, the corresponding ratio is about 8.5. However, there is no indication of high density clumps towards this area from high-resolution H {\scriptsize I} observations~\citep{park+13}, meaning that this scenario is unlikely. It is therefore possible that shell A is a separate SNR, or part of it, independent of CTB~80. 

\citet{mavromatakis+01} presented CCD images of CTB~80 in several optical lines and discovered a number of long and thin filaments in this area. They concluded that CTB~80 is more extended than earlier radio observations indicated. In the area of CTB~80 also Lynds Bright Nebula \citep[LBN,][]{lynds65} 156 and 158 are located, showing a complex filamentary structure in their surroundings, which complicates a clear identification of structure A. \citet{mavromatakis+02} found shock-heated filaments in the northeast of CTB~80, which they regarded as an indication for the existence of another SNR in this area, although its size and shape were not well constrained by the observed filaments. The two longest thin filaments, apparently emerging from LBN 156, run almost parallel to structure A, but at a distance of about 15$\arcmin$. These filaments have very faint radio counterparts. It seems thus possible, that shell A is part of the SNR proposed by \citet{mavromatakis+02}. 

\section{Conclusions}

We performed a polarization study of SNR CTB~80 using recent polarization observations at 1420~MHz, 2695~MHz, and 4800~MHz. We smoothed the $I$ and Stokes $Q$ and $U$ maps to a common resolution of $10\arcmin$, and derived maps of polarized intensity and polarization angle. Strong depolarization was found at 1420~MHz. We calculated RMs using polarization angles at 2695~MHz and 4800~MHz. The RM map shows a clear sign change towards shell C. The reason for this change is unclear. Combining the RM and depolarization maps, we found that the RM of CTB~80 contains a contribution from the thermal gas inside CTB~80 but not mixed with the emitting medium. We also identified a bright polarized structure with a very weak total-intensity correspondence, which could be either a part of CTB~80 or a part of another SNR independent of CTB~80. 

\begin{acknowledgements}
XL and XS are supported by the National Natural Science Foundation of China (Grant No. 11763008). XG is supported by the CAS-NWO cooperation programme (Grant No. GJHZ1865) and the National Natural Science Foundation of China (Grant No. U1831103). We thank Dr. Patricia Reich for reading of the manuscript and
discussions. We also thank the referee for critical comments that have improved the paper.
\end{acknowledgements}

\bibliographystyle{raa}
\bibliography{bibtex} 
\end{document}